# Short Range Ising Spin Glasses: a critical exponent study

E. Nogueira Jr.[1], S. Coutinho[2], F. D. Nobre[3], and E. M. F. Curado[4].

[1] *Instituto de Física,*
*Universidade Federal da Bahia*
*40210-340, Salvador, Bahia, Brazil.*

[2] *Laboratório de Física Teórica e Computacional,*
*Universidade Federal de Pernambuco,*
*50670-901, Recife, Pernambuco, Brazil.*

[3] *Departamento de Física Teórica e Experimental,*
*Universidade Federal do Rio Grande do Norte,*
*CP 1641, 59072-970, Natal, Rio Grande do Norte, Brazil.*

[4] *Centro Brasileiro de Pesquisas Físicas ,*
*Rua Xavier Sigaud, 150,*
*22290-180, Rio de Janeiro, Brazil.*

## Abstract

The critical properties of short-range Ising spin-glass models, defined on a diamond hierarchical lattice of graph fractal dimension $d_f = 2.58$, 3, and 4, and scaling factor 2 are studied via a method based on the Migdal-Kadanoff renormalization-group scheme. The order parameter critical exponent $\beta$ is directly estimated from the data of the local Edwards-Anderson (EA) order parameter, obtained through an exact recursion procedure. The scaling of the EA order parameter, leading to estimates of the $\nu$ exponent of the correlation length is also performed. Four distinct initial distributions of the quenched coupling constants (Gaussian, bimodal, uniform and exponential) are considered. Deviations from a universal behaviour are observed and analysed in the framework of the renormalized flow in a two dimensional appropriate parameter space.

Typeset using REVTEX

In this work, we report direct estimations of the critical temperature, the order parameter $\beta$ critical exponent and the correlation length critical exponent for the short range Ising spin glass models on the diamond hierarchical lattices (DHL) with graph fractal dimension $d_f = 2.58$, $3.0$ and $4.0$. By direct we mean that the exponents were computed from the numerical values of the local Edwards-Anderson (EA) order parameter obtained, through an exact recursive procedure developed by the authors to study the multifractal structure of the local order parameter [1]. This approach is based on the Migdal-Kadanoff renormalization group approximation for finite hypercubic lattices [2], which was proven to be exact, for pure systems, on the DHL. Within this method one is able to calculate the local magnetization of each site for each sample of the corresponding DHL as a function of the temperature. An scaling of the order parameter at $T = T_c$, makes it possible to obtain values for the exponent $\phi = \beta/\nu$, and therefore the correlation-length critical exponent $\nu$ as well.

The spin-glass Ising model on a diamond hierarchical lattice of arbitrary graph fractal dimension $d_f$, is defined by the hamiltonian

$$H = - \sum_{<i,j>} J_{ij} \sigma_i \sigma_j \qquad (1)$$

where $\{J_{ij}\}$ are the random quenched nearest-neighbor coupling constants following a given distribution and the $\sigma$'s represent the Ising variables assigned to the lattice sites. We consider four different distributions of coupling constants as an initial model condition, namely, *Gaussian, bimodal, uniform* and *exponential,* for lattices up to sixteen generations. The values of the Edwards-Anderson order parameter $q_i^{EA}$, defined for each lattice site, can be numerically evaluated for each initial distribution by means of an exact recursion procedure as a function of the temperature, for a general DHL with $N$ generations [1]. Adding such values over an appropriated set of lattice sites [1], for a given sample (one realization of the initial set of coupling constants) and taking the average over many samples, we end up with the usual EA order parameter per spin,

$$q^{EA} = \frac{1}{M} \left[ \sum_{i=1}^{M} q_i^{EA} \right]_J \qquad (2)$$

In the equation above, $[...]_J$ stands for the configurational average, whereas the sum is restricted to those lattices sites within a given shortest path joining the two end sites of the hierarchical lattice [1]; for a DHL with $N$ generations and scaling factor $b$, one has $M = b^N + 1$. Besides saving computer memory, this choice is expected to yield a stochastic representative set of local order parameters; after the average over many different samples is performed, any peculiarities of a particular path become irrelevant. The critical exponent $\beta$ is defined by

$$q^{EA} \sim [(T_c - T)/T_c]^\beta \quad (T \to T_c^-), \tag{3}$$

where $T_c$ is the spin-glass critical temperature, estimated by following numerically the probability distribution $P(t_{ij})$ associated with the thermal transmissivities, $\{t_{ij}\} \equiv \tanh(\beta\{J_{ij}\})$. For a centered initial distribution ($\langle t_{ij}\rangle = \langle J_{ij}\rangle = 0$), $T_c$ is defined as the temperature at which the width $\langle t_{ij}^2\rangle^{1/2}$ remains constant under the renormalization process [3]. Moreover, the correlation-length critical exponent $\nu$ can be also estimated by scaling the order parameter at the critical point, i.e.,

$$q^{EA} \sim M^{\beta/\nu} \quad (T = T_c). \tag{4}$$

Actually, we estimate the $\beta$ critical exponents for a $d_f = 2.58$, 3.0 and 4.0 DHL with scaling factor $b = 2$ by considering lattices with $N = 15$ generations and 300 samples. To obtain the ratio $\beta/\nu$ we consider $N$ varying from 8 to 15 and the number of samples varying from 1000 ( for smaller systems) to 300 (larger systems). The values of the local EA order parameters were calculated following the method developed by the authors [1], considering four distinct initial distributions of coupling constants namely, the *Gaussian, bimodal, exponential and uniform* distributions, all presenting an unitary width, given respectively by

$$P(J_{i,j}) = \tfrac{1}{\sqrt{2\pi}} \exp(-\tfrac{1}{2}J_{i,j}^2) \tag{5}$$

$$P(J_{i,j}) = \tfrac{1}{2}\left[\delta(J_{i,j} - 1) + \delta(J_{i,j} + 1)\right]$$

$$P(J_{i,j}) = \tfrac{1}{\sqrt{2}} \exp(-\sqrt{2}|J_{i,j}|)$$

$$P(J_{i,j}) = \begin{cases} \tfrac{1}{2\sqrt{3}} & if \ -\sqrt{3} \leq J_{i,j} \leq \sqrt{3} \\ 0 & \text{(otherwise)} \end{cases}$$

In Table I, we display our results for $\beta$ $\phi$ and $\nu$ critical exponents and critical temperatures for lattices with three graph fractal dimension $d_f$. It should be mentioned that as far as the hierarchical lattice is concerned, the only approximations involved in the present approach are the finite sizes, and the number of samples investigated.

For all lattice dimensionalities the critical temperatures were obtained by following the flow of the renormalized distribution width. In all cases the critical temperatures are smaller as higher is the kurtosis of the initial distribution of coupling constants. For lattices with graph fractal dimension $d_f = 2.58$, which correspond to a DHL basic unit with three connections and scaling factor 2, the critical temperatures are close but different from zero, indicating that the SG lower critical dimension $d_\ell$ is still lower than $d_f = 2.58$. Actually, within the MKRG approach $d_\ell$ was estimated to be around 2,43 [7,11].

The critical exponents $\beta$ and $\nu$ were obtained for all considered distributions except of the exponential one due to an avoided numerical overflow. Nevertheless, the figures obtained so far, shown an universal behavior within the error bars.

For lattices with $d_f = 3$ and 4, we have recovered recent results for the values of the critical temperatures obtained within the same approach. Moreover, whenever possible to compare, the agreement of our results with those obtained from other methods, developed for the cubic lattice (e.g., extensive Monte Carlo simulations [4,5], high temperature series expansions [6]) is very impressive. However, as far as the MKRG scheme is concerned there is a strong disagreement, in the Gaussian case, between our estimates of the exponent $\nu$ (1.8), and that obtained by Southern and Young (2.78) [3], evaluated also with a similar unit cell. The basic differences between their approach and ours are: (i) they force the bond distribution into a Gaussian at each step; (ii) they compute the $\nu$ exponent by monitoring the width of the distribution in the vicinity of the critical point, looking at the fluctuations around its fixed value, linearizing its recursion relation (linear approximation). It is a well-known fact [3,7] that the fixed-point distribution is indeed very close to the Gaussian one, so approximation (i) should not be too drastic. However the linear approximation has been criticized, and it may not be appropriate for the spin glasses, as it is for pure systems [8]; we suspect that this is the main reason for the disagreement between our estimate and the one of Ref. [3]. It is worth to mention that, for the case of a Gaussian distribution, the $\nu$ exponent calculated in the present work recovers the one obtained by McMillan [9], applying a method, called domain-wall renormalization group, based on the calculation of the domain-wall free energies for very small finite-size cubic lattices.

For $d_f = 3$, our results show that slight, but undeniable, different values of the exponent $\beta$ were obtained for the corresponding initial distributions of coupling constants, suggesting absence of universality in this model, as claimed by other authors [10]. However, this apparent lack of universality obtained within the present MKRG scheme should be interpreted carefully, taking into account the nature of the flow associated with the evolution of the coupling constants distributions, under the renormalization process. This is represented schematically in Fig. 1. If the renormalized distribution is forced to be shape-invariant at each renormalization step (as done in Ref. [3] for the Gaussian case), then the renormalized flow is restricted to a one-dimensional parameter subspace. However, if the distribution is let to evolve freely under the renormalization process, one may represent its flow in a higher dimensional parameter space. An illustrative two dimension projection of this space is the plane $\langle t_{ij}^2 \rangle^{1/2}$ versus $k_B T/\langle J_{ij}^2 \rangle^{1/2}$ [7,11]. Within this representation, the parameter space is separated in two regions, each one governed by its respective attractor (spin-glass and paramagnetic stable fixed points). In the frontier of these regions, one finds the saddle-point spin-glass critical point, characterized by a fixed-point distribution, which should remain unchangeable under renormalization. Since the analytical form of this distribution in unknown, the genuine critical properties associated with the spin-glass phase transition can not be determined through the present approach. Therefore, the results herein presented for each distribution, should be considered as "pseudo-critical", rather than critical properties. In Fig. 1 we also show typical flows starting from points slightly above and below the corresponding critical temperature of the bimodal distribution. Here, again, we notice that the position of the line $\langle t_{ij}^2 \rangle_p$ corresponding to each considered distribution also follows the decreasing of the kurtosis as we go from the left to the right in Fig. 1 [7,11]. In this work, the pseudo-critical $\beta$ exponents were obtained directly from the values of the local EA order parameters, calculated in the vicinity and below the critical temperatures associated with each considered probability distribution. For the present DHL, where $\frac{7}{8}$ of the total number of sites belong to the last generation, and whose local order parameters are calculated with the coupling constants introduced by the not-yet-renormalized (initial) distribution, different values of $\beta$ appear, describing the behavior of the system near these critical points; this should not be interpreted as a breakdown of universality. To obtain the genuine critical exponents, one should probe the critical region around the saddle-point spin-glass critical point a task which deserves a careful numerical analysis. We notice that for higher space dimensions (i.e., large number of parallel branches in the basic unit cell), the genuine fixed-point distribution should approach the Gaussian one [12].

In higher dimensions ($d_f = 4$) the $\beta$ exponent approaches to the mean field value for spin glasses with slight differences arising from the consideration of distinct initial distribution of couplings as above discussed for the three dimensional case.

Finally, we have also estimated the $\nu$ exponent by calculating the scaling exponent $\phi = \beta/\nu$ at the critical temperature corresponding to each distribution. We have obtained the same value for the correlation-length exponent, within the error bars, for the considered distributions. Moreover, these values are in good agreement with the corresponding values obtained by high-temperature series expansions and numerical simulations for the spin-glass Ising model on a three dimension cubic lattice [4,9,5] . It is amazing how the estimates for the spin-glass critical temperatures and exponents, on a diamond hierarchical lattice with scaling factor $b = 2$ and graph fractal dimension $d_f = 3$, can be so close to those of the corresponding Bravais lattice; why this happens for complex systems like spin glasses, and not to simple ferromagnets, is a point which is now deserving further studies [11]. For lower and higher dimensions both exponents are slightly independent from the initial distribution do not suggesting breaking of universality.

## ACKNOWLEDGMENTS

This research was supported by CNPq, FINEP and CAPES. One of us (SC) is also grateful to CNPq for individual financial support under Grant AIP 523757/94-8

TABLES

| Dimension | | Bimodal | Uniform | Gaussian | Uniform |
|---|---|---|---|---|---|
| 2.58 | $T_c$ | 0.48 | 0.35 | 0.29 | 0.22 |
| | $\beta$ | 0.25 ±0.02 | 0.28 ±0.02 | 0.28±0.02 | |
| | $\beta/\nu$ | 0.119 ±0.002 | 0.122±0.002 | 0.128±0.004 | |
| | $\nu$ | 2.1±0.1 | 2.3±0.2 | 2.3±0.3 | |
| 3 | $T_c$ | 1.15 | 0.96 | 0.88 | 0.75 |
| | $\beta$ | 0.73 ±0.06 | 0.44 ±0.05 | 0.63 ±0.06 | 0.91±0.08 |
| | $\beta/\nu$ | 0.43 | 0.26 | 0.35 | 0.49 |
| | $\nu$ | 1.7 ±0.1 | 1.7 ±0.2 | 1.8 ±0.2 | 1.9 ±0.2 |
| 4 | $T_c$ | 2.31 | 2.20 | 2.08 | 1.88 |
| | $\beta$ | 0.99 ±0.01 | 0.87 ±0.01 | 0.99±0.01 | 0.91 ±0.01 |
| | $\beta/\nu$ | | | | |
| | $\nu$ | 1.51 ±0.02 | 1.32 ±0.02 | 1.30 ±0.01 | 1.52 ±0.01 |

TABLE I. The estimated values for the critical temperature $T_c$ and critical exponents $\beta$, $\beta/\nu$ and $\nu$. The columns are distributed accordingly with the increasing of the initial distribution kurtosis. Note that the critical temperatures decreases as the kurtosis increases.

FIGURES

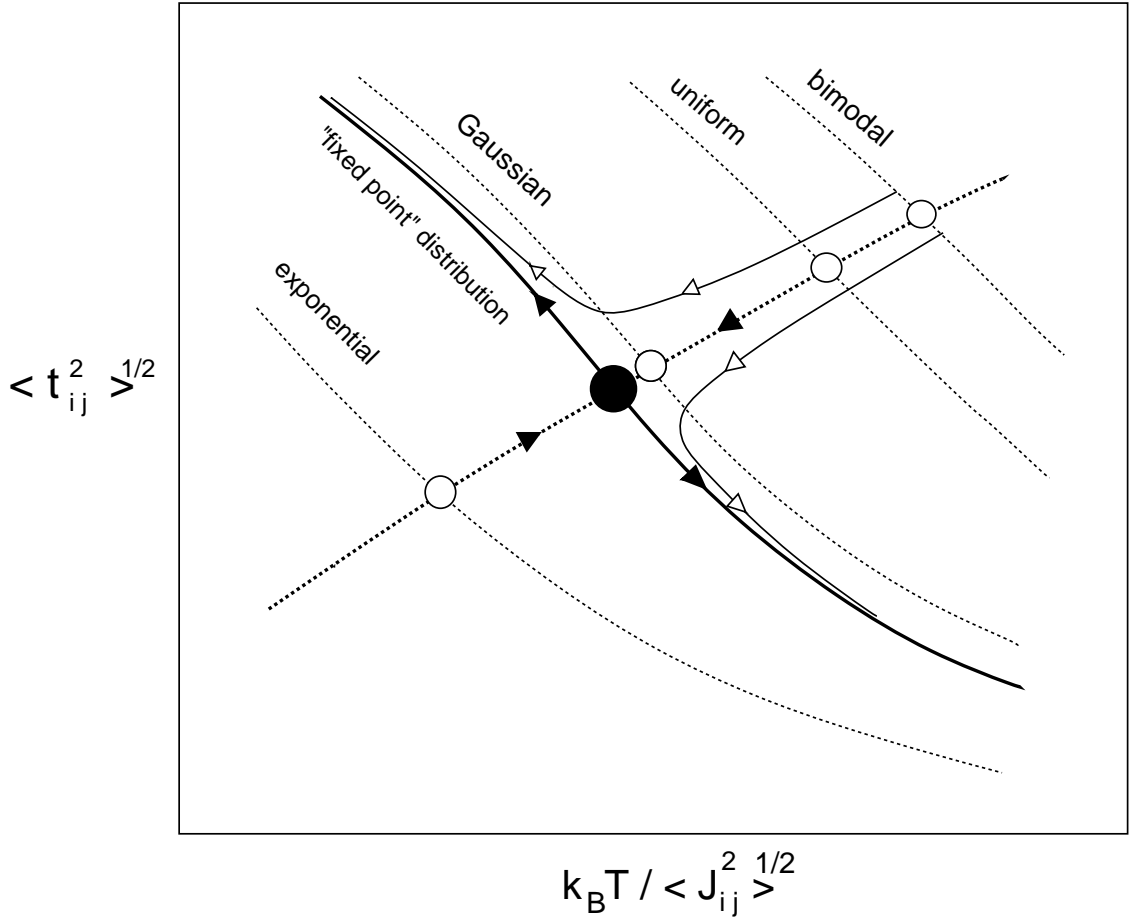

FIG. 1. Schematic flow diagram in the appropriated parameter space for the Ising SG model on the DHL. The full thick line represents the "fixed-point distribution", whereas the dashed lines, the non-renormalized (initial) distributions: exponential, Gaussian, uniform and bimodal. Full lines indicate the flow starting from a hypothetical point, close to critical temperature of the bimodal initial distribution. The full circle represents the spin-glass fixed critical point and open circles, the critical points corresponding to each of the initial distributions

9